\documentclass[conference]{IEEEtran}
\IEEEoverridecommandlockouts
\usepackage{cite}
\usepackage{amsmath,amssymb,amsfonts}
\usepackage{algorithm, algorithmic}
\usepackage{graphicx, subcaption}
\usepackage{balance}
\usepackage{textcomp}
\usepackage{xcolor}
\usepackage{hyperref}

\def\BibTeX{{\rm B\kern-.05em{\sc i\kern-.025em b}\kern-.08em
    T\kern-.1667em\lower.7ex\hbox{E}\kern-.125emX}}
\renewcommand{\baselinestretch}{0.975}

\begin{document}
\bstctlcite{IEEEexample:BSTcontrol}
\title{RIS Optimization Algorithms for Urban Wireless Scenarios in Sionna RT}
\author{\IEEEauthorblockN{
Ahmet Esad Güneşer\textsuperscript{$\ast\circ$}, 
Berkay Şekeroğlu\textsuperscript{$\ast\bullet$},
Sefa Kayraklık\textsuperscript{$\ast$},
Erhan Karakoca\textsuperscript{$\ast$},\\
İbrahim Hökelek\textsuperscript{$\ast\diamond$},
Sultan Aldirmaz-Colak\textsuperscript{$\circ$},
Ali Görçin\textsuperscript{$\ast\diamond$}
}
\IEEEauthorblockA{\textsuperscript{$\ast$}\href{https://hisar.bilgem.tubitak.gov.tr/en/}{Communications and Signal Processing Research (HISAR) Lab.}, TUBITAK BILGEM, Kocaeli, Türkiye\\
\textsuperscript{$\circ$}Department of Electronics and Communication Engineering, Kocaeli University, Kocaeli, Türkiye\\
\textsuperscript{$\bullet$}Department of Electrical and Electronics Engineering, Boğaziçi University, Istanbul, Türkiye\\
\textsuperscript{$\diamond$}Department of Electronics and Telecommunications Engineering Istanbul Technical University, Istanbul, Türkiye\\
		Email:
		aesadguneser@hotmail.com,
        berkay.sekeroglu@tubitak.gov.tr,
        sefa.kayraklik@tubitak.gov.tr,\\
        erhan.karakoca@tubitak.gov.tr,
        ibrahim.hokelek@tubitak.gov.tr,
        sultan.aldirmaz@kocaeli.edu.tr,
        ali.gorcin@tubitak.gov.tr
		\vspace{-3ex}}
}

\maketitle
\begin{abstract}
This paper evaluates the performance of reconfigurable intelligent surface (RIS) optimization algorithms, which utilize channel estimation methods, in ray tracing (RT) simulations within urban digital twin environments. Beyond Sionna's native capabilities, we implement and benchmark additional RIS optimization algorithms based on channel estimation, enabling an evaluation of RIS strategies under various deployment conditions. Coverage maps for RIS-assisted communication systems are generated through the integration of Sionna's RT simulations. 
Moreover, real-world experimentation underscores the necessity of validating algorithms in near-realistic simulation environments, as minor variations in measurement setups can significantly affect performance.
\end{abstract}


\section{Introduction}
The rapid evolution of wireless communication systems has introduced unprecedented challenges in meeting the demands of higher data rates, enhanced coverage, and energy efficiency, particularly in dense urban environments \cite{chowdhury20206g}. As cities become increasingly interconnected and data-driven, new paradigms such as reconfigurable intelligent surfaces (RIS) have emerged as transformative technologies for optimizing wireless networks \cite{basar2019wireless}. By intelligently manipulating electromagnetic waves, RIS has the potential to enhance signal propagation, reduce interference, and improve network performance in challenging urban scenarios. Simultaneously, the concept of urban digital twins offers a powerful framework for simulating, analyzing, and optimizing next-generation communication systems. This paper bridges these two emerging technologies, leveraging urban digital twin environments and ray tracing (RT) simulations to evaluate and optimize RIS-assisted communication systems, laying a foundation for designing more efficient and adaptive wireless networks.

RIS is a strong candidate for the sixth-generation cellular system. RIS has been investigated in depth in the literature. To demonstrate the performance of RIS-enabled systems, both RT-based simulations and measurements on real testbeds have been performed \cite{huang2022novel, huang2023ray, m_amri_1_sparsity_aware, hao2024modeling}.
In \cite{huang2022novel}, the received power distribution in the indoor environment is obtained using a RT simulator at 5.4 GHz for two cases with and without RIS to demonstrate the effect of RIS on the received signal. In this study, simulations were performed for various cases, such as the location of the RIS and whether there is a line-of-sight (LoS) between the transmitter and the receiver. 
The authors in \cite{huang2023ray} proposed a deterministic channel model based on RT for RIS-assisted wireless communication and obtained channel impulse response performing data processing. Both performance gain and reduction in delay spread and spread angle through the RIS implementation were verified through a comparison of simulation and measurement results.    
A real RIS testbed with 512 passive elements with 1-bit resolution is used in \cite{m_amri_1_sparsity_aware}. Sparsity-aware channel estimation is performed based on compressive sensing (CS) algorithms, such as the Dantzig selector and orthogonal matching pursuit (OMP). The authors in \cite{m_amri_1_sparsity_aware} conducted experiments both in indoor and outdoor environments. The Hadamard method and the mirror-like surface (MLS), where all unit cells reflect the incident wave with 0$^{\circ}$ phase shift, were used as references to compare the performance of the proposed system. The authors showed that the OMP method utilizes merely one-fourth of the training samples necessary for the Hadamard method. It has been shown that with the proposed technique, signals are received with 15 dBm and 7 dBm higher power than MLS when the distances between the RIS-Rx are 9 meters and 33 meters for an outdoor experiment, respectively.
Another RIS measurement and RT simulation are realized in \cite{hao2024modeling}. In this study, the received power is tabulated for the cases where each element of the RIS has $\{1, 2, 3, 4\}$-bit resolution and continuous resolution. It was demonstrated that the proposed RIS model from an electromagnetic perspective aligns with the developed communication models. 
Furthermore, in \cite{guo2023wireless}, a hybrid sparse channel estimation method utilizing wireless beacons for RIS-aided mmWave communications is presented.

Tensor-based open-source library Sionna\cite{sionna} has emerged as a wireless physical layer simulation environment, including RT capabilities for radio wave propagation. Sionna allows rapid implementations of complex communication architectures, such as end-to-end RIS-assisted wireless communication systems. In \cite{shabanpour2024physically}, the authors optimized RIS reradiation modes using Sionna RT, designed unit cells and validated their approach through fabrication and experiments. Recently, Sionna has been integrated with the network simulator ns-3 in order to provide an open-source full-stack digital network twin \cite{pegurri2024toward} and to employ a RT capability \cite{zubow2024ns3}. 

By leveraging the Sionna simulation framework, we utilize a RT model to analyze the effectiveness of different RIS configurations in realistic urban scenarios.
Beyond the native capabilities of the Sionna framework, we implement and benchmark several RIS optimization algorithms, which are inspired by the work of \cite{m_amri_1_sparsity_aware}, that are not originally included in Sionna. This makes it possible to assess RIS strategies more broadly under various deployment scenarios.
We also generated coverage maps for RIS-assisted communication systems. These maps provide valuable insights into the spatial impact of RIS optimization on coverage enhancement in urban areas. This integration highlights the transformative potential of digital twins as a powerful tool for modeling and improving RIS-assisted communication systems in complex urban settings.
Furthermore, real-world experimentation demonstrates that before deploying the developed algorithms, it is crucial to validate them in a near-realistic simulation environment since minor variations in the measurement setup can lead to performance fluctuations. 

The organization of this paper is as follows: The system mode is introduced in Section II. Channel estimation and RIS configuration are explained in Section III. Simulation and measurement results are given in Section IV. Section V concludes the paper.

\section{System Model}\label{sec:system_model}
\subsection{Channel Model}
A wireless communication system comprising a single-antenna transmitter, a single-antenna receiver, and a passive RIS is considered. The passive RIS, which has a size of $K \times L$ is divided into unit cells arranged in $K$ rows and $L$ columns.
The received signal can be expressed as
 \begin{equation}
    y = h x + n,
\end{equation}
where $h$ represents end-to-end channel, $x$ is the pilot signal transmitted by the transmitter, and $y$ is the signal received at the receiver. The additive white Gaussian noise (AWGN) is denoted as $n$.

Considering all possible paths, the end-to-end channel gain $h$ can be decomposed into the non-line-of-sight (NLOS) and line-of-sight (LOS) components. For the NLOS path, the received signal is observed as the summation of all signals reflected from each RIS unit cell, where ${h}^\text{Tx-RIS-Rx}_{k,l}$ denotes the gain of the channel involving the unit cell $(k,l)$,
\begin{equation} \label{eq:endtoend1}
    h = \underbrace{h^\text{Tx-RIS-Rx}}_{\text{NLOS}} + \underbrace{h^\text{Tx-Rx}}_{\text{LOS}} = \sum^K_{k=1} \sum^L_{l=1} h^\text{Tx-RIS-Rx}_{k,l} + h^\text{Tx-Rx}.
\end{equation}
We can decompose $h^\text{Tx-RIS-Rx}_{k,l}$, by seperately modeling the channel between Tx-RIS ($h^\text{Tx-RIS}_{k,l}$), and RIS-Rx ($h^\text{RIS-Rx}_{k,l}$) as 
\begin{equation} 
    \label{eq:decompose}
    h^\text{Tx-RIS-Rx}_{k,l} = h^\text{Tx-RIS}_{k,l} \Phi^{}_{k,l} h^\text{RIS-Rx}_{k,l},
\end{equation}
where $\Phi_{k,l}$ denotes the reflection coefficient of the RIS unit cell $(k, l)$. Therefore, substituting \eqref{eq:decompose} into \eqref{eq:endtoend1}, the end-to-end channel gain $h$ can be expressed as,
\begin{equation}
\begin{aligned}
h = & \sum^K_{k=1} \sum^L_{l=1} h^\text{Tx-RIS-Rx}_{k,l} + h^\text{Tx-Rx} \\
    = & \sum^K_{k=1} \sum^L_{l=1} h^\text{Tx-RIS}_{k,l} \Phi_{k,l} h^\text{RIS-Rx}_{k,l} + h^\text{Tx-Rx}.
\end{aligned}
\end{equation}
The end-to-end channel gain, excluding the RIS reflection coefficient, is defined as the RIS channel gain:
\begin{equation}
\label{gmn}
\varrho_{k,l} = h^\text{Tx-RIS}_{k,l} h^\text{RIS-Rx}_{k,l},
\end{equation}
where $ h^\text{Tx-RIS}_{k,l}$ and $h^\text{RIS-Rx}_{k,l}$  represent the channel gains from the Tx to the RIS and from the RIS to the Rx, respectively. The objective is to estimate $\varrho_{k,l}$, enabling the configuration of the RIS reflection coefficients $\Phi_{k,l}$ to maximize the overall channel gain.
\subsection{Simulating the Channel with Ray Tracing}
In \textit{Sionna} \cite{sionna}, RT is used to compute propagation paths between transmitters and receivers, including both NLOS and LOS components. Once the channel impulse response is generated with \colorbox{gray!20}{\texttt{cir()}}, the channel frequency response is obtained using \colorbox{gray!20}{\texttt{cir\_to\_ofdm\_channel()}}, providing the channel coefficients \(h\). Repeating this process for each path computed through RT yields the coefficients \(h^\text{Tx-RIS-Rx}_{k,l}\).
\begin{figure}[t!]
    \centering
    \includegraphics[width=1\linewidth]{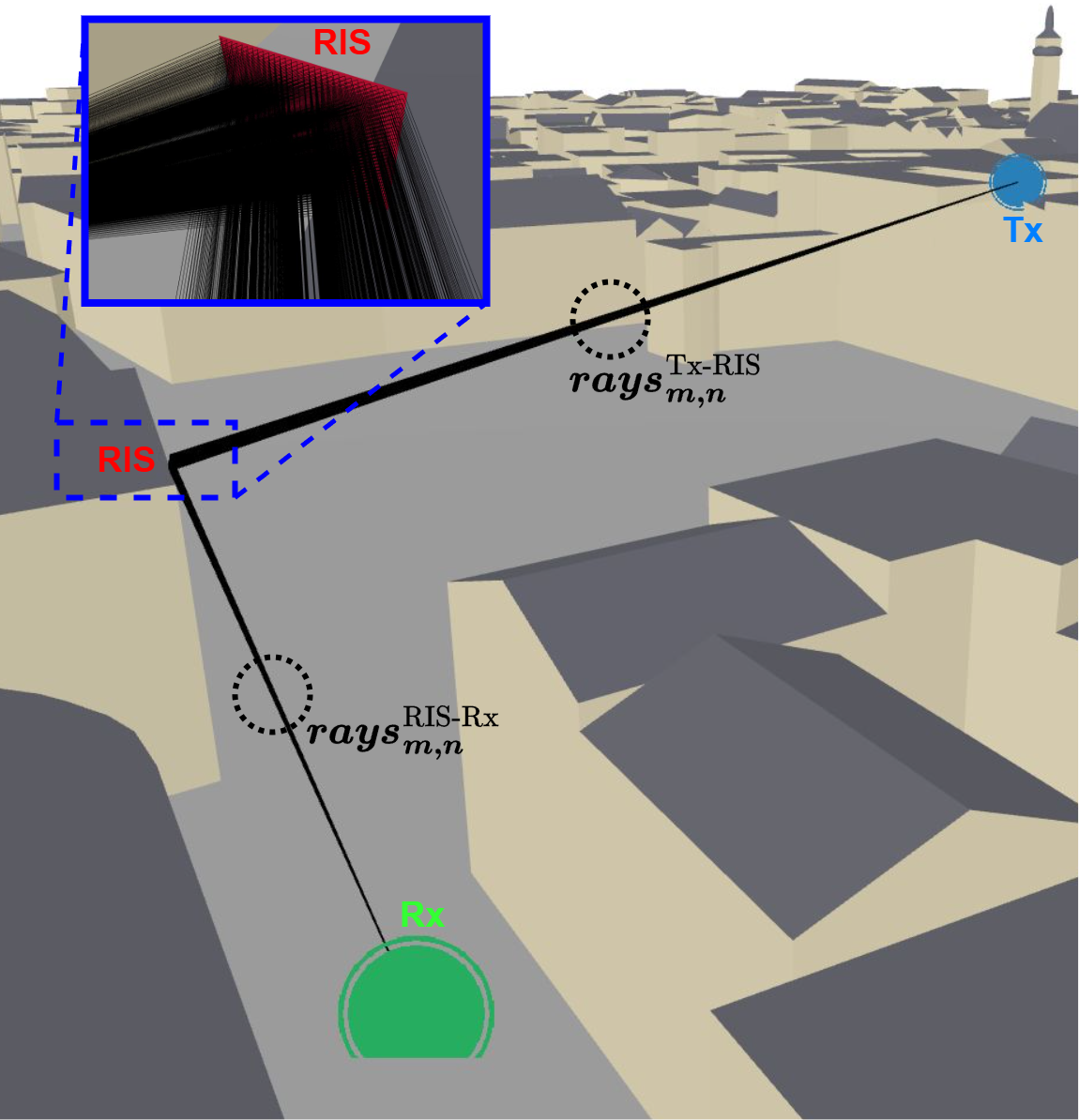}
    \caption{RT-based RIS paths computed for Munich city scenario.}
    \label{fig:sionnaScene}
    \vspace{-15pt}
\end{figure}

The RIS reflection coefficient in \textit{Sionna} is defined by an amplitude and a phase profiles, jointly represented as $\boldsymbol{\Phi}$, which determines the reradiated electromagnetic field. Both profiles are \(K \times L\) arrays. By default, the amplitude profile is an array of ones, while the phase profile is an array of zeros. Under these default settings, each RIS unit cell behaves as an identity element (\(\Phi_{k,l} = 1\)), and the computed \(h^\text{Tx-RIS-Rx}_{k,l}\) in 
\eqref{eq:decompose} corresponds to $h^\text{Tx-RIS}_{k,l} h^\text{RIS-Rx}_{k,l}$, as given in 
~\eqref{gmn}. Thus, with the default RIS configuration, \(\varrho_{k,l}\) can be directly obtained using \colorbox{gray!20}{\texttt{cir\_to\_ofdm\_channel()}}.

In a practical experimental setup, only the received signal \( y\) would be observable, providing a noisy estimation of \( h^\text{Tx-RIS-Rx} \), the end-to-end channel gain (summed over all unit cells). The objective is to maximize \( h^\text{Tx-RIS-Rx} \) by adjusting the RIS configuration (\( \Phi_{k,l} \)). Accurate estimation of \( \varrho_{k,l} \), the channel gain without the RIS contribution, enables setting \( \Phi_{k,l} \) optimally to maximize the overall end-to-end channel gain.

\section{Channel Estimation and RIS Configuration Methods}
Through a series of configurations, where the \(m^\text{th}\) RIS configuration is represented as \(\mathbf{\overline{\Phi}}_m\) (a vector of dimension \(KL\)), a known pilot signal $\mathbf{x}$ is transmitted, and the received signal $\mathbf{y}$ is observed, assuming the channel is sparse. Let \(\mathbf{\overline{\Phi}}_m\) and \(\boldsymbol{\overline{\varrho}}\) denote the flattened vectors for the RIS configuration matrix \(\mathbf{\Phi}\) and the RIS channel gain matrix \(\boldsymbol{\varrho}\), respectively. For the \(m^\text{th}\) configuration, the noisy estimation of the end-to-end channel is denoted as \(\widehat{{h}}_m\), while $\widehat{{n}}_m$ represents the estimation noise for the same configuration. In a setup excluding the LOS link, the received signal for m-th configuration is expressed as:
\begin{equation}
    {\widehat{h}}_m = \sum_{k=1}^K \sum_{l=1}^L \varrho_{k,l} \Phi_{k,l} + \widehat{n}_m = \mathbf{\overline{\Phi}}_m^{T} \boldsymbol{\overline{\varrho}} + {\widehat{n}}_m.
\end{equation}
Repeating this process for different configurations (\(m = 1, 2, \dots, M\)) provides noisy estimations of the end-to-end channel for all configurations, forming a vector \(\widehat{\mathbf{h}}\) of dimension \(M\). By arranging the vectors \(\mathbf{\overline{\Phi}}_M^{T}\) from each configuration into the rows of a \(M \times KL\) matrix, the sensing matrix \(\mathbf{W}\) is constructed. Using \(\widehat{\mathbf{h}} = (\widehat{h}_1, \ldots, \widehat{h}_M)^T\) and \(\widehat{\mathbf{n}} = (\widehat{n}_1, \ldots, \widehat{{n}}_M)^T\), the channel estimation for all configurations can be written as:
\begin{equation}
    \widehat{\mathbf{h}} = \mathbf{W} \boldsymbol{\overline{\varrho}} + \widehat{\mathbf{n}}.
\end{equation}
Under the sparsity condition of \(\boldsymbol{\overline{\varrho}}\), which can be represented in the angular domain, the following relationship holds:
\begin{equation}
    \boldsymbol{\Gamma} = \frac{1}{KL} \mathbf{U}^* \boldsymbol{\overline{\varrho}},
\end{equation}
where \(\mathbf{U}\) is $KL\times KL$ dimensional two-dimensional discrete Fourier transform matrix mapping the channel gain to its angular representation. The sensed vector for the RIS configurations can then be expressed as:
\begin{equation}
\begin{aligned}
    \widehat{\mathbf{h}} = \mathbf{W} \boldsymbol{\overline{\varrho}} + \widehat{\mathbf{n}} = \mathbf{W} \mathbf{U}^* \boldsymbol{\Gamma} + \widehat{\mathbf{n}}.
\end{aligned}
\end{equation}
The channel estimation algorithms aim to design $\mathbf{W}$ where it can reveal sparsity characteristics of the channel and an algorithm to estimate $\boldsymbol{\varrho}$.

After estimating the channel coefficients, the RIS can compensate the channel by applying phase shift configurations.
\begin{align} \label{eq:conjChan}
    \boldsymbol{\psi} = -\angle{\boldsymbol{{\varrho}}} 
\end{align}
$\boldsymbol{\psi}$ is the conjugate phase of the channel coefficient, a vector of size \textit{KL}. A wireless channel has a continuous phase range from 0° to 180°, but RIS hardware only supports binary phase shifts - either 0° or 180°. Therefore, the estimated channel phase must be quantized to these values before being applied to the RIS elements, resulting in performance degradation due to loss of continuous phase resolution.
\begin{equation}
    \boldsymbol{\Bar{\psi}} =
    \begin{cases}
        \pi , & \frac{\pi}{2} \leq \boldsymbol{\psi} \leq \frac{3\pi}{2} \\
        0, & \text{otherwise}
    \end{cases}\label{eq:quantized}
\end{equation}
In \eqref{eq:quantized}, the obtained phases are quantized according to the RIS hardware. $\boldsymbol{\Bar{\psi}}$ contains the quantized most appropriate phase values, the column vector of size \textit{KL}. The obtained values should be used as new reflection coefficients of the RIS hardware.

\begin{figure*}[t]
\centering
\subfloat[]{\label{fig:simConf:a} 
\includegraphics[width=0.255\linewidth]{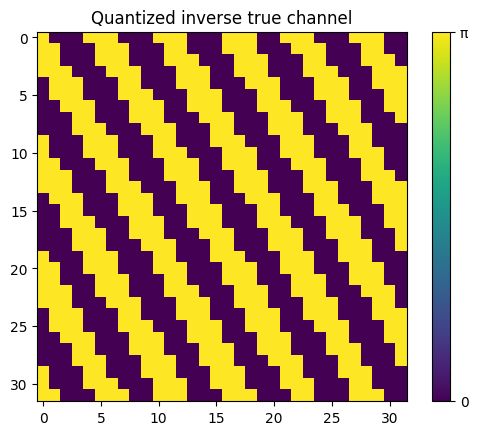}}
\hspace{-0.5cm}
\subfloat[]{\label{fig:simConf:b} 
\includegraphics[width=0.255\linewidth]{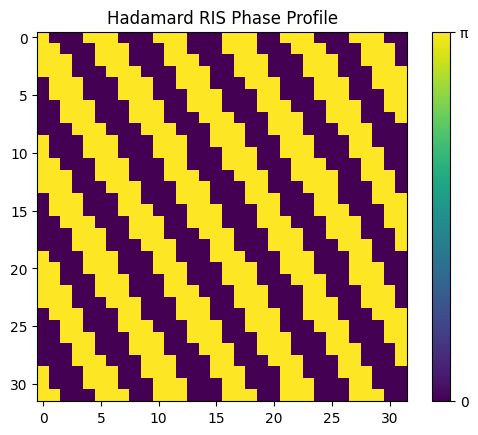}}
\hspace{-0.5cm}
\subfloat[]{\label{fig:simConf:c} 
\includegraphics[width=0.255\linewidth]{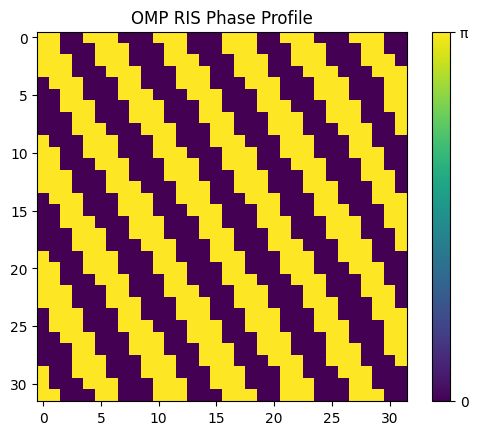}}
\hspace{-0.5cm}
\subfloat[]{\label{fig:simConf:d} 
\includegraphics[width=0.255\linewidth]{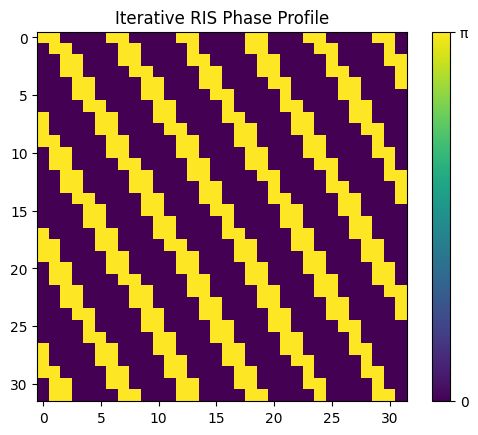}}
\caption{The RIS of size \( 32 \times 32\)  configurations for the cases of the true channel, Hadamard, OMP, and iterative methods, respectively.}
\label{fig:simConf}
\vspace{-12pt}
\end{figure*}

\begin{figure*}[t]
\centering
\subfloat[]{\label{fig:simCov:a} 
\includegraphics[width=0.245\linewidth]{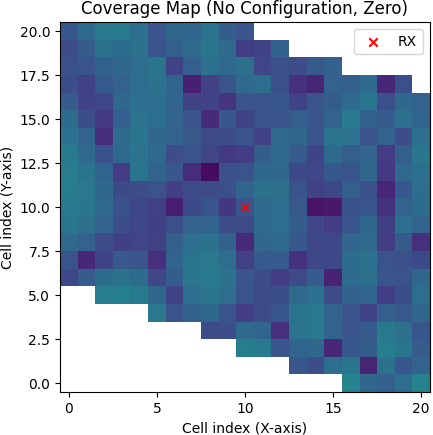}}
\subfloat[]{\label{fig:simCov:b} 
\includegraphics[width=0.218\linewidth]{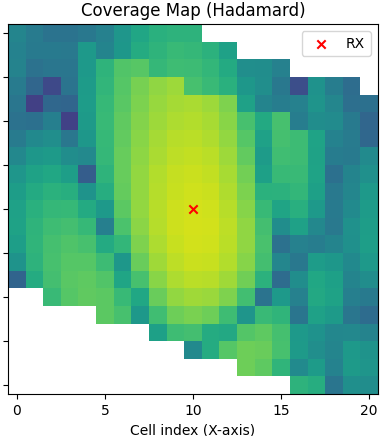}}
\subfloat[]{\label{fig:simCov:c} 
\includegraphics[width=0.215\linewidth]{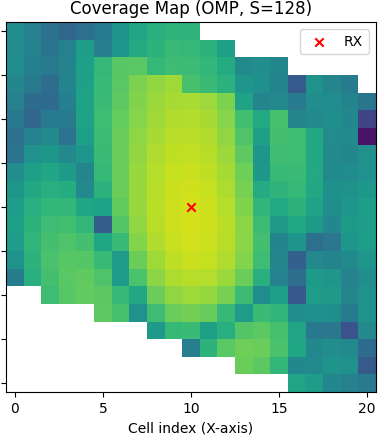}}
\subfloat[]{\label{fig:simCov:d} 
\includegraphics[width=0.278\linewidth]{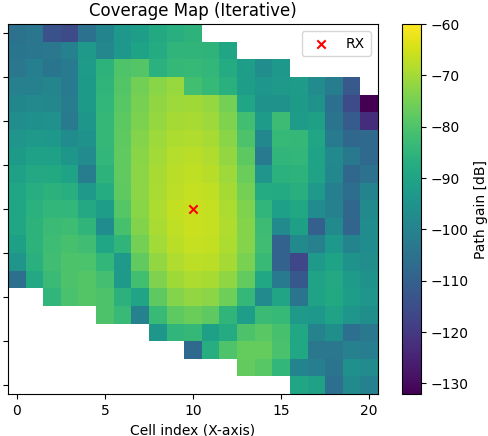}}
\caption{The signal coverage for the cases of RIS-off (RIS phase profile set to 0), Hadamard, OMP, and iterative methods, respectively (Axes are in meters).}
\label{fig:simCov}
\vspace{-15pt}
\end{figure*}

\subsection{Hadamard Matrix Based Channel Estimation}
The sensing matrix \(\mathbf{W}\) can be chosen as a Hadamard matrix of size \( KL \times KL\) where $M=KL$, denoted as \(\mathbf{\Sigma}_{KL}\). A Hadamard matrix is a square matrix with elements subset of $\{+1,-1\}$, and its rows and columns are orthogonal. The size of the matrix must be a power of 2 (\(2^b\)). The Hadamard matrix is recursively defined as follows:
\begin{equation}
\begin{split}
         \mathbf{\Sigma}_{2^b} &= \begin{bmatrix}
        \mathbf{\Sigma}_{2^{b-1}} & \mathbf{\Sigma}_{2^{b-1}} \\
        \mathbf{\Sigma}_{2^{b-1}} & -\mathbf{\Sigma}_{2^{b-1}}
    \end{bmatrix}.
\end{split}
\label{eq:hadamard}
\end{equation}
Using the rows of \(\mathbf{W}\) as RIS configurations, the RIS channel gain vector \(\boldsymbol{\overline{\varrho}}\) can be estimated by multiplying the observed channel vector \(\widehat{\mathbf{h}}\) with the transpose of the sensing matrix, \(\mathbf{W}^T\). The Hadamard matrix satisfies the orthogonality property $\mathbf{\Sigma}_{KL}^T \mathbf{\Sigma}_{KL} = KL \mathbf{I}$, where \(\mathbf{I}\) is the identity matrix of size \( KL \times KL\). Under noise-free conditions, this ensures that the estimate of \(\boldsymbol{\overline{\varrho}}\) equals the true channel gain.

By setting \(\mathbf{W} = \mathbf{\Sigma}_{KL}\), the least-squares (LS) estimate of \(\mathbf{\overline{\varrho}}\) can be expressed as:
\begin{equation}
    \widehat{\mathbf{h}} = \mathbf{W} \boldsymbol{\overline{\varrho}} + \widehat{\mathbf{n}},
\xrightarrow{\text{LS estimate } \boldsymbol{\overline{\varrho}}} \,\, \widehat{\boldsymbol{\overline{\varrho}}} = \frac{\mathbf{W}^T \widehat{\mathbf{h}}}{KL} 
    = \boldsymbol{\overline{\varrho}} + \frac{\mathbf{W}^T \widehat{\mathbf{n}}}{KL},
\end{equation}
where \(\widehat{\mathbf{h}}\) is the vector of noisy channel observations, \(\widehat{\mathbf{n}}\) is the noise vector, and \(\widehat{\boldsymbol{\overline{\varrho}}}\) is the LS estimate of the RIS channel gain vector \(\boldsymbol{\overline{\varrho}}\).
The steps for channel estimation using the Hadamard sensing matrix are summarized in Algorithm \ref{alg:hadamard} \cite{m_amri_1_sparsity_aware}.
\begin{algorithm}[ht!]
\caption{Hadamard Channel Estimation Algorithm} \label{alg:hadamard}
\begin{algorithmic}[1]
\renewcommand{\algorithmicrequire}{\textbf{Input:}}
\renewcommand{\algorithmicensure}{\textbf{Output:}}
\REQUIRE Pilot signal \(x\), sensing matrix \(\mathbf{W}\), number of RIS rows (\(K\)) and columns (\(L\)).
\ENSURE RIS channel gain vector \(\boldsymbol{\overline{\varrho}}\)
\FOR{$i = 1$ to $KL$}
    \STATE Load the \(i^\text{th}\) row of \(\mathbf{W}\) (\(\mathbf{W}_{(i,:)}\)) to the RIS configuration.
    \STATE Transmit the pilot signal \(x\) and measure the received signal \(y\).
    \STATE Compute the channel observation: \(\widehat{h}_i = y / x\).
\ENDFOR
\STATE Compute the RIS channel: \(\boldsymbol{\overline{\varrho}} = \frac{\mathbf{W}^T \widehat{\mathbf{h}}}{KL}\).
\RETURN \(\boldsymbol{\overline{\varrho}}\)
\end{algorithmic}
\end{algorithm}

\subsection{Orthogonal Matching Pursuit Based Channel Estimation}
In the CS framework, the number of rows in the sensing matrix rows $(M)$ can be much smaller than the total number of RIS unit cells (\(KL\)), provided that the channel representation \(\boldsymbol{\Gamma}\) exhibits sparsity. A randomized approach based on the Bernoulli distribution is employed to construct the sensing matrix \(\mathbf{W}\) for CS. Specifically, each entry in \(\mathbf{W}\) is independently selected as either \(+1\) or \(-1\) with equal probability. The sparse angular domain channel vector \(\boldsymbol{\Gamma}\) can be recovered using the OMP algorithm, which is applied for RIS channel estimation. It is worth noting that, while OMP estimates \(\boldsymbol{\Gamma}\), the spatial domain channel vector \(\boldsymbol{\varrho}\) can be obtained by simply applying an inverse discrete Fourier transform to \(\boldsymbol{\Gamma}\). The steps for OMP-based channel estimation are detailed in Algorithm \ref{alg:omp} \cite{m_amri_1_sparsity_aware}.

\begin{algorithm}[ht!]
\caption{Orthogonal Matching Pursuit} \label{alg:omp}
\begin{algorithmic}[1]
\renewcommand{\algorithmicrequire}{\textbf{Input:}}
\renewcommand{\algorithmicensure}{\textbf{Output:}}
\REQUIRE Columns of \(\mathbf{W} \mathbf{U}^*\) (\(\mathbf{v}_j\) denotes the \(j^\text{th}\) column of \(\mathbf{W} \mathbf{U}^*\)),  
         sensing result vector \(\widehat{\mathbf{h}}\),  
         sparsity level \(S\)
\ENSURE Indices of nonzero entries \(\gamma\),  
         values of nonzero entries \(\boldsymbol{\Gamma}\)
\STATE \textbf{Initialization:}  
\(\gamma \gets \emptyset\) (empty vector),  
\(\mathbf{\Omega} \gets \emptyset\) (empty matrix),  
\(\mathbf{r} \gets \widehat{\mathbf{h}}\) (initialize residual)

\FOR{\(k = 1\) to \(S\)}
    \STATE \(j^* \gets \operatorname{argmax}_{j=1,\dots,KL}\frac{\left|\mathbf{v}_j^H \mathbf{r}\right|^2}{\mathbf{v}_j^H \mathbf{v}_j}\)
    \STATE Append \(\mathbf{v}_{j^*}\) to \(\mathbf{\Omega}\): \(\mathbf{\Omega} \gets \left[\mathbf{\Omega}, \mathbf{v}_{j^*}\right]\)
    \STATE Append \(j^*\) to \(\gamma\): \(\gamma \gets \gamma \cup \{j^*\}\)
    \STATE Compute the sparse channel:  
    \(\boldsymbol{\Gamma} \gets \left(\mathbf{\Omega}^H \mathbf{\Omega}\right)^{-1} \mathbf{\Omega}^H \widehat{\mathbf{h}}\)
    \STATE Update the residual:  
    \(\mathbf{r} \gets \widehat{\mathbf{h}} - \mathbf{\Omega} \boldsymbol{\Gamma}\)
\ENDFOR

\RETURN \(\gamma, \boldsymbol{\Gamma}\)
\end{algorithmic}
\end{algorithm}

\subsection{Iterative Method}
The RIS phase shift configuration for focusing the incoming beams toward the receiver is computed by adjusting the phase shifts of the reflecting elements and measuring the received signal power during each iteration. Since evaluating all possible RIS configurations is computationally impractical, an iterative method \cite{iterative} is employed to find an RIS configuration for reflecting signals in the desired direction, avoiding the need for exhaustive searching. Firstly, all reflecting elements are set to no phase shift mode. Then, the possible states of the first element are applied, and the state that maximizes the received signal power is selected. After the first element's best state is found, this process is repeated separately for each remaining element. Note that the iterative method optimizes the RIS phase shift configuration by inspecting the received signal power instead of estimating the RIS channel coefficients.

\section{Simulation and Measurement Results}

\subsection{Simulations and Coverage Map}
The results of the simulation scenarios are presented in this section. Cartesian coordinates \((x, y, z)\) are used to define the locations of the transmitter (Tx), reconfigurable intelligent surface (RIS), and receiver (Rx). The positions are as follows: Tx at \([23.79, 16.54, 28]\), RIS centered at \([0, 0, 0]\), and Rx at \([-23, 156, 2]\). The dimensions of the RIS surface $K\times L$ are set to 32$\times$32 for simulations, with a spacing of 0.5$\lambda$ between unit cells where $\lambda$ denotes wavelength. The system operates at a frequency of $5.2$ GHz. Each method requires broadcast instants for pilot transmissions. Hadamard and OMP use these pilots to measure the channel for the RIS configuration \(\mathbf{W}_{(i,:)}\), while the iterative method relies on the received signal strength to determine the on/off state of each unit cell. The broadcast instants are set to 1024 for Hadamard, 512 for OMP, and 1024 for the iterative method. The Munich is used as a RT scenario in Sionna. All the simulations were performed in Google Colab using T4 GPU backed.

The simulation results highlight the impact of various RIS configuration methods on the phase profiles and their corresponding coverage maps. The quantized inverse true channel Fig. \ref{fig:simConf:a}, representing the ground truth under binary phase constraints, serves as the reference for comparison. The Hadamard phase profile has a structured binary pattern that nearly matches the real channel, which is shown in Fig. \ref{fig:simConf:b}. In addition, the OMP phase profile in Fig. \ref{fig:simConf:c} shows a binary structure that closely resembles the ground truth, demonstrating the OMP algorithm's capacity to properly reconstruct the channel under sparsity assumption while balancing complexity and performance. Even though the iterative phase profile is not similar to the quantized inverse true channel in Fig. \ref{fig:simConf:d}, it exhibits a diagonal structure shaped by the optimization process. It is important to note that iterative methods generally require greater computational resources than approaches such as Hadamard or OMP, making them less suitable for real-time applications despite their potential performance benefits.

The corresponding coverage maps further illustrate the impact of these configurations on system performance. Without specific RIS configuration (unit cells are set to zeros and act as a reflective metal plate), as shown in Fig. \ref{fig:simCov:a}, the signal strength is weak and randomly dispersed, with no clear focus around the receiver, at around -90 dB. This demonstrates the importance of RIS configuration. The Hadamard-based configuration in Fig. \ref{fig:simCov:b} achieves a well-concentrated signal around the receiver, with strong path gains close to the center. Likewise, the OMP configuration Fig. \ref{fig:simCov:c} achieves focused path gains around the receiver, comparable to the Hadamard method. OMP is also effective in sparse channel environments, providing reliable performance with reasonable computational requirements. In addition, the iterative method also delivers a similar coverage map shown in Fig. \ref{fig:simCov:d}. For all three methods, a path gain of -60 dB is observed at the center, representing a 30 dB improvement compared to the RIS-off scenario. However, this method typically requires higher computational resources than Hadamard or OMP. Nevertheless, in urban environments, OMP's and Hadamard's sparse assumption may falter due to increased multipath, leading to performance loss. However, moderate non-sparsity still permits acceptable results. Meanwhile, iterative RIS configuration, adapting based on real-time received signal power improvements without assuming sparsity, remains robust in such challenging conditions.
\subsection{Computation Times}
The computation times of the three RIS configuration methods vary due to differences in design and processing. Hadamard and OMP rely on channel estimation, requiring pilot signals, with Hadamard needing as many broadcast instants as RIS unit cells and OMP reducing this by half. Their computation times are $0.019$ s and $1.219$ s, respectively, excluding pilot transmissions and focusing on algorithm processing. In contrast, the iterative method $25.861$ s  requires broadcast instants equal to RIS unit cells and recalculates the channel gain for each transmission, leading to significantly higher computation time. Iterative approaches must be executed sequentially to accurately track performance improvements, which prevents the use of parallel processing or hardware acceleration. Thus, computation time increases with the number of RIS elements.
\begin{figure}[t]
    \centering 
    \includegraphics[width=0.9\linewidth]{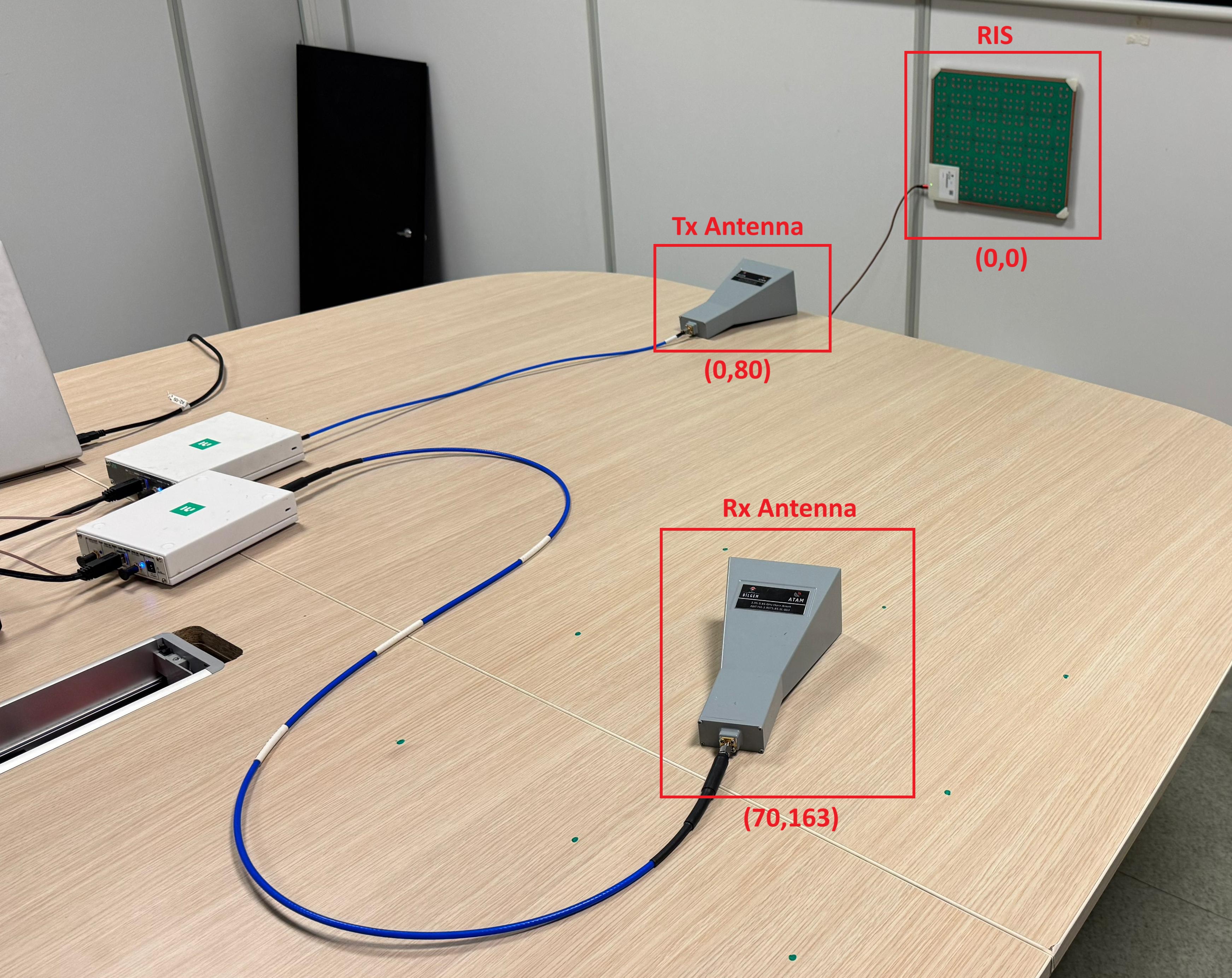}
    \caption{The measurement setup of the RIS-aided communication system.}
    \label{fig:measSetup} 
    \vspace{-15pt}
\end{figure}
\vspace{-5pt}
\subsection{Measurement Results}
The measurement setup of the RIS-assisted wireless communication system is illustrated in Fig. \ref{fig:measSetup}. The over-the-air signals, which are generated using MATLAB, are transmitted and received at the center frequency of $5.2$ GHz by utilizing USRP B210 software-defined radios (SDRs). The Greenerwave RIS prototype \cite{greenerwave} is fixed to the wall to steer the incoming signal toward the receiver, where both the horn antennas of the transmitter and the receiver are directed to the RIS. The reflecting unit elements of $8 \times 10$ with a uniform planar array design comprise the RIS, excluding its $2\times 2$ lower left corner for its controller. The phase shifts of the reflecting $76$ unit element on the RIS can be independently controlled using PIN diodes as $0^\circ$ and $180^\circ$. 
\begin{figure}[t]
    \centering 
    \includegraphics[width=0.85\linewidth]{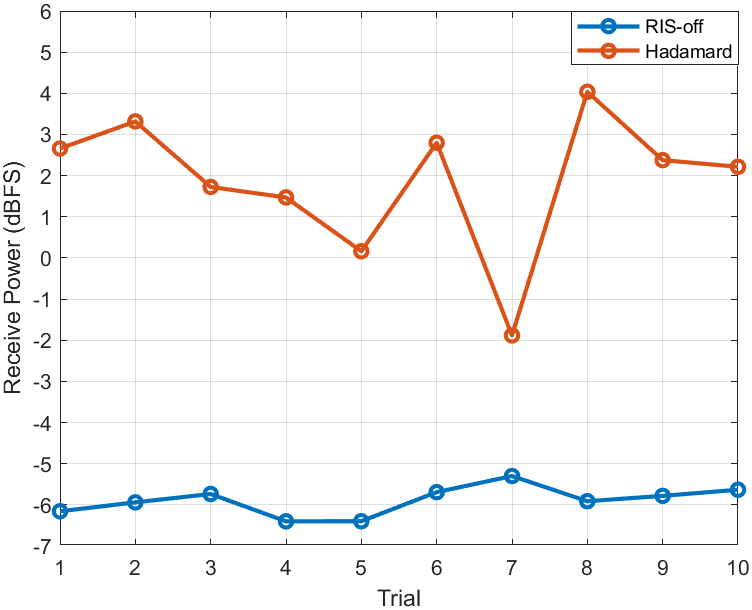}
    \caption{The received signal power when the RIS is configured using the Hadamard approach for 10 separate measurements.}
     \label{fig:measResult}
     \vspace{-20pt}
\end{figure}

The phase shift configuration of the RIS is determined according to the Hadamard channel estimation method, which is described in Algorithm \ref{alg:hadamard}. The channel estimation algorithm computes the channel corresponding to each reflecting unit element of the RIS by applying the rows of the Hadamard matrix. Due to the structure of the Hadamard matrix, the size of the $80 \times 80$ matrix is generated, and only the corresponding $76$ rows and columns are utilized in Algorithm \ref{alg:hadamard} since the RIS prototype consists of $76$ reflecting elements. Note that the channel observation through the iterations of the algorithm is conducted by modifying the example receiver code in the MathWorks website \cite{mathworks}. After estimating the channels for the reflecting elements of the RIS, the RIS configuration is found by first taking the conjugate phase of the channel and quantizing the corresponding phase shift into $0^\circ$ and $180^\circ$ according to (\ref{eq:conjChan}) and (\ref{eq:quantized}), respectively.  

Throughout the experiment process, $10$ separate trials, each of which includes more than $400$ measurements, are conducted by configuring the RIS phase shifts according to the channel estimations from Algorithm \ref{alg:hadamard}. The average received signal powers of the measurements for each trial are illustrated in Fig. \ref{fig:measResult} for the cases when the RIS is turned off and configured according to the Hadamard channel estimation. It can be observed from Fig. \ref{fig:measResult} that the RIS can increase the received signal power on average around $7$ dB except for the $7$th trial. The variability observed across different trials in Fig. \ref{fig:measResult} can be attributed to hardware and practical limitations of the devices used in the experiments. Since the Hadamard algorithm relies on accurate channel estimation, even minor variations introduced by the transceiver hardware can degrade the algorithm's performance. Therefore, prior to employing the developed algorithms in real-world experimentation, it is essential to validate them in a near-realistic simulation environment such as Sionna's RT capabilities. However, differences between the results arise due to the $32 \times 32$ RIS size and Urban scenario used in the simulation environment, compared to the $8 \times 10$ RIS size and indoor scenario used in real-time measurements.

\section{Conclusion}
This paper examines the performance of RIS through RT simulations in Sionna and real-world measurements. Hadamard and OMP algorithms are applied to optimize RIS phase configurations based on channel estimation. Although RT is employed here for channel modeling, the study's broader objective is to advance digital twin frameworks by investigating time-varying channel dynamics and RIS optimization strategies. This work establishes a foundation for dynamic and scalable network optimization within digital twins, supporting more precise wireless system designs in rapidly changing urban environments. Additionally, the measurement results reveal that integrating an RIS with one-bit resolution into the wireless communication system improves the received signal strength by at least 3 dB and up to 9 dB compared to scenarios without RIS.

\bibliographystyle{IEEEtran}
\bibliography{main}
\end{document}